\documentclass[]{interact}

\usepackage{epstopdf}
\usepackage[caption=false]{subfig}

\usepackage[numbers,sort&compress]{natbib}
\bibpunct[, ]{[}{]}{,}{n}{,}{,}
\makeatletter
\def\NAT@def@citea{\def\@citea{\NAT@separator}}
\makeatother

\theoremstyle{plain}

\theoremstyle{definition}

\theoremstyle{remark}

\usepackage{tikz}
\usepackage{color, soul}
\usepackage{overpic}
\usepackage{amsmath}
\usepackage{comment}
\newcommand{\beq}{\begin{equation}}
\newcommand{\eeq}{\end{equation}}
\newcommand{\un}[1]{\,\mathrm{#1}}
\newcommand{\lb}{\left(}
\newcommand{\rb}{\right)}
\begin{document}


\title{Turbulent shear layers in confining channels}

\author{
\name{Graham P. Benham\textsuperscript{a}\thanks{CONTACT Graham P. Benham. Email: benham@maths.ox.ac.uk},
  Alfonso A. Castrejon-Pita\textsuperscript{b},  Ian J. Hewitt\textsuperscript{a}, Colin P. Please\textsuperscript{a}, Rob W. Style\textsuperscript{c}  and Paul A. D. Bird\textsuperscript{d}}
\affil{\textsuperscript{a}Mathematical Institute, University of Oxford, Andrew Wiles Building, Radcliffe Observatory Quarter, Woodstock Road, Oxford OX2 6GG United Kingdom; 
\textsuperscript{b}Department of Engineering Science, University of Oxford, Oxford OX1 3PS, United Kingdom;
\textsuperscript{c}Department of Materials, ETH Z\"{u}rich, Z\"{u}rich 8093, Switzerland;
\textsuperscript{d} VerdErg Renewable Energy Limited, 6 Old London Rd, Kingston upon Thames KT2 6QF, United Kingdom;
}
}

\maketitle

\begin{abstract}
We present a simple model for the development of shear layers between parallel flows in confining channels.
Such flows are important across a wide range of topics from diffusers, nozzles and ducts to urban air flow and geophysical fluid dynamics.
The model approximates the flow in the shear layer as a linear profile separating uniform-velocity streams. Both the channel geometry and wall drag affect the development of the flow.
The model shows good agreement with both particle image velocimetry experiments and computational turbulence modelling. {The simplicity and} low computational cost of the model allows it to be used for {benchmark predictions and} design purposes, which we demonstrate by investigating optimal pressure recovery in diffusers with non-uniform inflow.
\end{abstract}


\section{Introduction}
Shear layers, where two parallel flows undergo turbulent mixing, are found in a wide variety of situations.
For laterally unconfined flows, shear layers can be described using a simple analytical model - derived from the turbulent boundary layer equations and Prandtl mixing length theory \citep{schlichting1960boundary, jimenez2004turbulence}.
However, this model is often not practicable in many situations where the flows are confined in a channel.
Examples of such flows include 
blockages and cavities in pipes \citep{griffith2007wake, bailey2013pressure, suresha2016nonlinear}, 
air flow in urban environments \citep{sau2003interaction, alam2011two}, 
environmental flows \citep{michelsen2015two, venditti2014flow, anderson2014numerical}, 
engine and aerodynamic design \citep{simone2012analysis, chamorro2013interaction, kang2014onset}, 
and mixing flows in diffusers, nozzles and pumps \citep{lo2012separation, zhang2011erosion, shah2013experimental}.
In such cases, it is important to account for the confining effect of the channel walls.

{In addition to the large literature on unconfined shear layers, as detailed in the review articles by Brown and Roshko} \citep{brown2012turbulent} {or Huerre} \cite{huerre2000open}, {there are some numerical and experimental studies of confined shear layers.
For example, Gathmann et al. }\citep{gathmann1993numerical} {use numerical simulations to investigate the transition to turbulence in a confined shear layer. Furthermore, the criteria for absolute and convective instabilities in a confined compressible shear layer are studied by Robinet et al. }\citep{robinet2001wall}. {Further numerical studies of confined shear layers are discussed in} \cite{tam1989instability,sau2003interaction, lavalle2017ultraefficient}. 
{In the context of experimental investigations of confined shear layers, Bradshaw et al.} \citep{bradshaw1972reattachment} {studied the behaviour of a shear layer near a backward-facing step. In addition, Castro et al.} \citep{castro1987structure} {calculated the shear layer structure in a separation region near a wall. The use of confined shear layers for vector thrusting is investigated by Alvi et al.} \citep{alvi2000vectoring}. 
{
Other examples of experimental investigations of confined shear layers are given by} \citep{lander2016influence, alam2011two, suresha2016nonlinear, chu1988confinement}.
{However, while there are a limited number of experimental and numerical studies, there is no simple model for confined shear layers.}

Here we show that a simple model, comprising two uniform streams separated by a linear shear layer and incorporating parameterisation of wall drag, can be used to model confined shear layers {and very good agreement is attained with} both experiments and detailed $k$-$\epsilon$ turbulence modelling. 
{The decomposition of the flow profile into uniform streams and a linear shear layer is similar to that discussed by Jimenez} {\cite{jimenez2004turbulence}}{, except that we account for confinement effects, since the velocities of the plug regions are non-constant. Furthermore, our model accounts for the interaction between the shear layer and the channel walls. The evolution of the shear profile is governed by an equation which we derive from an entrainment argument, and which is also analogous to the classic analytical model for the growth of unconfined shear layers} {\cite{schlichting1960boundary, jimenez2004turbulence}}. 

The simplicity of our model means it both gives good physical intuition into the flow, and is {much cheaper than Computational Fluid Dynamics (CFD) models.  Furthermore, it can be applied to shear layers in a channel of arbitrary shape.} Thus it can be used to quickly find optimal parameters in a wide range of engineering design problems \citep{shah2013experimental}.
It also avoids the need for high levels of expertise that are typically required with CFD when choosing turbulence models and selecting boundary conditions \citep{schlichting1960boundary, pope2000turbulent}.
{As an example, we use our simple model to find the optimum design of a flow diffuser with a confined shear layer, where the objective is to maximise the pressure recovery. The results of the optimisation show that the optimum diffuser shape strikes a balance between not widening to soon, which would accentuate the non-uniform flow, and not being narrow for too long, which is detrimental for drag.  }

\section{Mathematical model\label{model}}
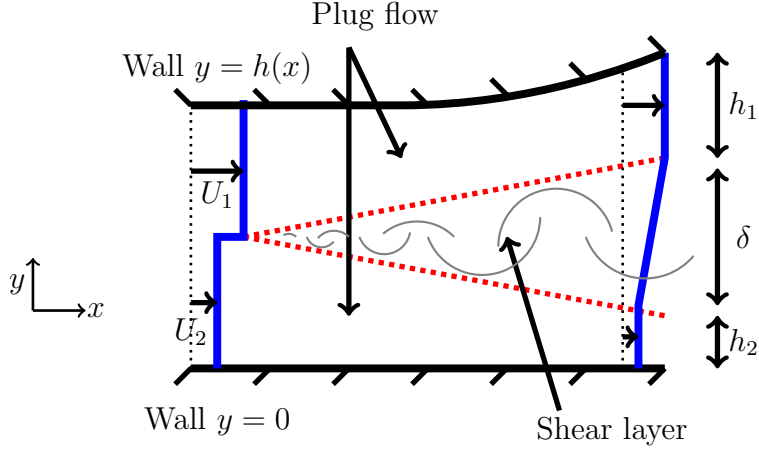
\begin{figure}
\centering
\begin{tikzpicture}[scale=0.7]
\draw[line width=3] (1,0) -- (10,0); 
\draw[red,dotted,line width=2] (2,5/2) -- (10,4); 
\draw[red,dotted,line width=2] (2,5/2) -- (10,1); 
\draw[blue, line width=3] (3/2, 0) -- (3/2,5/2) -- (2,5/2) -- (2,5.1) ; 
\draw[blue, line width=3] (9.5, 0) -- (9.5,1.2) -- (10,4) -- (10,6) ; 
\draw[->,line width=2] (1,15/4) -- (2,15/4); 
\draw[->,line width=2] (1,5/4) -- (3/2,5/4); 
\draw[->,line width=2] (9.2,1.2/2) -- (9.5,1.2/2); 
\draw[->,line width=2] (9.2,5) -- (10,5); 
\draw[dotted,line width=1] (1,0) -- (1,5); 
\draw[dotted,line width=1] (9.2,0) -- (9.2,52/9); 
\draw[->,line width=2]  (8,-0.8) --(7,5/2); 
\draw[->,line width=2] (4,6.1) -- (5,4); 
\draw[->,line width=2] (4,6.1) -- (4,1); 
\node at (4.5,6.7) {\large Plug flow};
\node at (9,-1.2) {\large Shear layer};
\draw [thick,gray] (3,2.5) arc [radius=0.2, start angle=45, end angle= 120];
\draw [thick,gray] (3.2,2.5) arc [radius=0.3, start angle=200, end angle= 320];
\draw [thick,gray] (4.0,2.5) arc [radius=0.4, start angle=45, end angle= 140];
\draw [thick,gray] (4.2,2.5) arc [radius=0.5, start angle=200, end angle= 350];
\draw [thick,gray] (6.0,2.5) arc [radius=1.0, start angle=45, end angle= 130];
\draw [thick,gray] (5.5,2.5) arc [radius=1.1, start angle=200, end angle= 360];
\draw [thick,gray] (9,2.6) arc [radius=1.1, start angle=15, end angle= 180];
\draw [thick,gray] (8.5,2.5) arc [radius=1.2, start angle=200, end angle= 320];
\draw[<->,line width=2] (11,4) -- (11,6);
\draw[<->,line width=2] (11,1.2) -- (11,3.8);
\draw[<->,line width=2] (11,0) -- (11,1);
\node at (11.5,5) {\large $h_1$};
\node at (11.5,2.5) {\large $\delta$};
\node at (11.5,0.5) {\large $h_2$};
\node at (1.5,3.3) {\large $U_1$};
\node at (1,0.7) {\large $U_2$};
\draw[black,line width=3,-] (1,5) parabola bend (5,5) (10,6);
\draw[line width=2] (1,5) -- (0.7,5.3); 
\draw[line width=2] (2.5,5) -- (2.2,5.3); 
\draw[line width=2] (4,5) -- (3.7,5.3); 
\draw[line width=2] (5.5,5) -- (5.2,5.3); 
\draw[line width=2] (7,5.2) -- (6.7,5.5); 
\draw[line width=2] (8.5,5.5) -- (8.2,5.8); 
\draw[line width=2] (10,6) -- (9.7,6.3); 
\draw[line width=2] (1,0) -- (0.7,-0.3); 
\draw[line width=2] (2.5,0) -- (2.2,-0.3); 
\draw[line width=2] (4,0) -- (3.7,-0.3); 
\draw[line width=2] (5.5,0) -- (5.2,-0.3); 
\draw[line width=2] (7,0) -- (6.7,-0.3); 
\draw[line width=2] (8.5,0) -- (8.2,-0.3); 
\draw[line width=2] (10,0) -- (9.7,-0.3); 
\draw[->,line width=1] (-2,1.1) -- (-1,1.1);
\draw[->,line width=1] (-2,1.1) -- (-2,2.1);
\node at (-0.8,1.1) {\large $x$};
\node at (-2.3,1.6) {\large $y$};
\node at (1.5, 5.7) {\large Wall $y=h(x)$};
\node at (1.5,-1) {\large Wall $y=0$};
\end{tikzpicture}
\caption{Schematic diagram of the velocity profile in a channel. The flow can be divided into three regions: Two plug flow regions of different speeds (in which Bernoulli's equation holds) and a turbulent shear layer between them.
{We do not resolve boundary layers at the walls, but instead we parameterise their effect using a friction factor $f$.}\label{tikz}}
\end{figure}

The flow geometry we consider is shown in Figure \ref{tikz}, {in which we illustrate our chosen coordinate system $(x,y)$.}
We consider two-dimensional turbulent flow in a long, thin channel where the rate of change of the channel width is small. 
To model this, we generalise the classical model for a shear layer between unconfined flows \citep{schlichting1960boundary,jimenez2004turbulence}, where a flow with velocity $U_1$ in the $x$ direction meets a second, parallel flow with velocity $U_2< U_1$.
A shear layer forms at the point they meet and grows with a width $\delta(x)$.
To good approximation, the average velocity in the shear layer increases linearly from $U_2$ to $U_1$ across its width \citep{jimenez2004turbulence}. Furthermore from the turbulent boundary layer equations and Prandtl's mixing length theory, it can be shown \citep{schlichting1960boundary, jimenez2004turbulence} that
\beq
\frac{d\delta}{dx}=2S\frac{U_1-U_2}{U_1+U_2},\label{delta}
\eeq
where $S$ is the spreading parameter, which is a non-dimensional number that has been determined by numerous free shear layer experiments {giving $S=0.06-0.11$} \citep{pope2000turbulent}.

{It is not clear that free shear layer theory should apply to confined situations, nor whether we expect the same value of the spreading parameter $S$. However, we incorporate a modified version of (\ref{delta}) in our model and find that it shows good comparison with both experimental and CFD calculations. }
In particular, when the flow is confined, the shear layer may reach the channel walls, so that its width can no longer evolve according to (\ref{delta}). To accommodate this we reinterpret (\ref{delta}) as describing how the shear in the layer behaves rather than how the width develops. To do this we introduce the gradient of the velocity profile, or shear rate, $\epsilon_y = (U_1-U_2)/\delta$. {Using this new variable, (\ref{delta}) is rewritten as
\beq
\frac{U_1+U_2}{2}\frac{d \epsilon_y}{dx}=-S_c\epsilon_y^2,\label{shear}
\eeq
where we denote $S_c$ as the spreading coefficient associated with confined shear layers.} We make the key assumption that the evolution of the shear rate continues according to (\ref{shear}) even when the shear layer is adjacent to the wall. A useful interpretation of this equation is that, moving along the channel at the average velocity in the shear layer $(U_1 + U_2)/2$, the shear rate decays at a rate proportional to the square of itself. 
{Equation (\ref{shear}) can also be derived from an entrainment argument, as described in Appendix \ref{appA}.} {The parameter $S_c$, like $S$, must be determined from experiments.
}

For the case of a confined flow we combine this description of the shear layer with conservation equations to describe the plug-like flows on either side, whose speeds $U_1$ and $U_2$ now vary with $x$. 
{Such equations are not necessary for a free shear layer when $U_1$ and $U_2$ are constant.} 
We consider flow in a channel $0<y<h(x)$, as shown schematically in Figure \ref{tikz}. Coflowing streams mix inside the channel. The velocity profile is approximated as
\begin{equation}
{
u=\begin{cases}
U_{2} &: 0<y<h_2,\\
U_2+\epsilon_y\lb y-h_2\rb &: h_2<y<h-h_1,\\
U_{1} &: h-h_1<y<h,
\end{cases}\label{piecewise}
}
\end{equation}
where $h_1(x)$ and $h_2(x)$ vary with distance along the channel. The width of the shear layer is $\delta=h-h_1-h_2$ and is related to the shear rate $\epsilon_y$ by definition $\delta=(U_1-U_2)/\epsilon_y$. 
{In this model we do not resolve the boundary layers at the wall, but instead we parameterise their effect using a friction factor (discussed later). A more complex model could include boundary layer growth, but we find that this approach compares well with experiments and CFD, and is attractive for its simplicity.} 
Additionally, assuming that the channel is much longer than it is wide, from boundary layer theory \citep{schlichting1960boundary}, the pressure is approximately uniform across the channel width and $p=p(x)$. 
By invoking conservation of mass and momentum, we can now predict how $U_1,U_2,h_1,h_2$ and $p$ evolve.
Averaged across the channel, the mass and momentum equations are
\begin{align}
 \int_0^h \rho u\, dy& = Q,\label{mass}\\
\quad \frac{d}{d x}\lb  \int_0^h  \rho u^2\, dy \rb+h\frac{d p}{d x}&=\tau_{w_0}+\tau_{w_h},
\label{mass_moment}
\end{align}
where $\rho$ is the density and $Q$ is the constant mass flux. $\tau_{w_0}$ and  $\tau_{w_h}$ are the wall stresses associated with walls at $y=0$ and $y=h$, respectively. 
{We parameterise the wall stress terms with a friction factor $f$ such that $\tau_{w_h}=-\frac{1}{8}f\rho U_1^2$ and $\tau_{w_0}=-\frac{1}{8}f\rho U_2^2$.
To estimate $f$ we use the empirical Blasius relationship \citep{blasius1913ahnlichkeitsgesetz, mckeon2005new} for flow in smooth pipes $f=0.316/Re^{1/4}$. 
Although this expression was derived for fully-developed pipe flow, we use it here for partially developed flows throughout the geometries we consider because we find good comparison with CFD and experiments. Further discussion of other approaches to estimate $f$ is addressed later.

}
{We assume that the main contributions to energy dissipation come from the wall drag and turbulent fluctuations in the shear layer. We ignore the energy dissipation in the unmixed plug flow regions since it is small by comparison \citep{schlichting1960boundary}.} Thus, in the plug regions we assume Bernoulli's equation \citep{batchelor2000introduction} holds {along streamlines, ignoring transverse velocity components since they are small, such that}
\beq
p+\frac{1}{2}\rho U_i^2=\frac{1}{2}\rho U_{i}(0)^2 \quad i=1,2,\label{bernoulli}
\eeq
where we take $p(0)=0$ as the reference pressure. 
Now, for a given channel shape $h(x)$,  and inlet conditions $U_1(0)$,  $U_2(0)$, $h_1(0)$ and $h_2(0)$, we can solve the system of differential algebraic equations (\ref{shear})-(\ref{bernoulli}) to find
$u(x,y)$ and $p(x)$.

\section{Comparison with $k$-$\epsilon$ model and experiments}

\begin{figure}
\centering
\subfloat{
\begin{overpic}[width=0.45\textwidth]{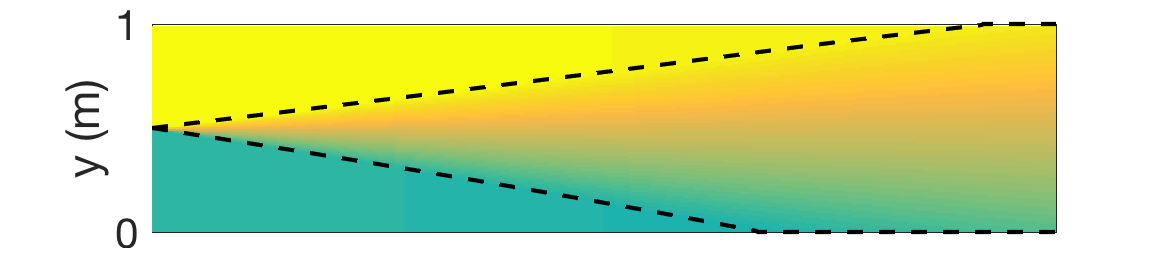}
\put(-5,20){(a)}
\put(35,22){\bf Simple model}
\end{overpic}
\begin{overpic}[width=0.45\textwidth]{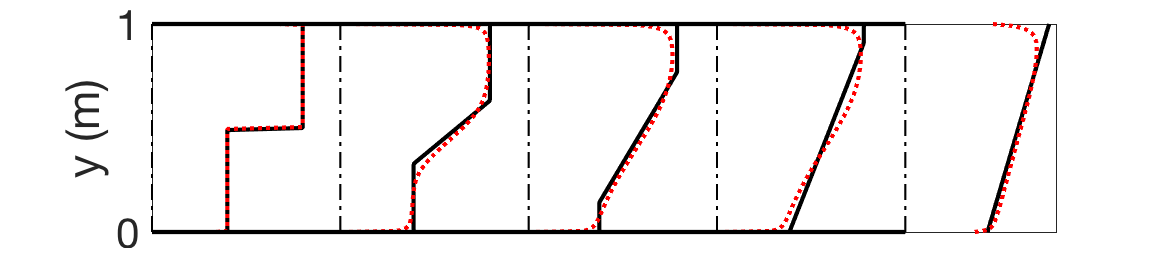}
\put(-5,20){(c)}
\end{overpic}
}
\vspace{0.2cm}\\
\subfloat{
\begin{overpic}[width=0.45\textwidth]{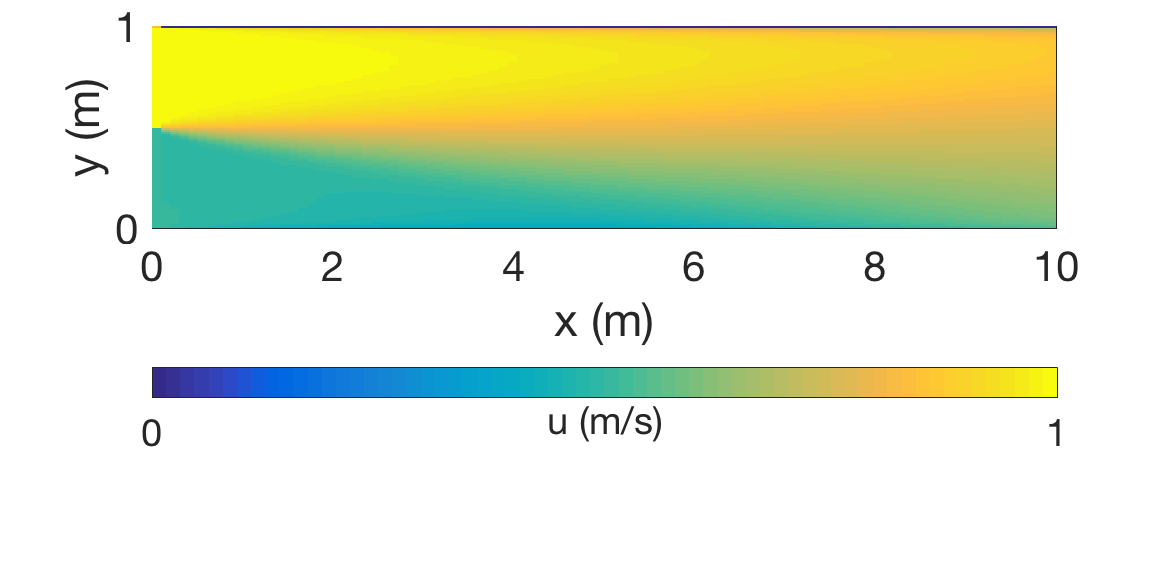}
\put(-5,45){(b)}
\put(40,49){\bf $k$-$\epsilon$ model}
\end{overpic}
\begin{overpic}[width=0.45\textwidth]{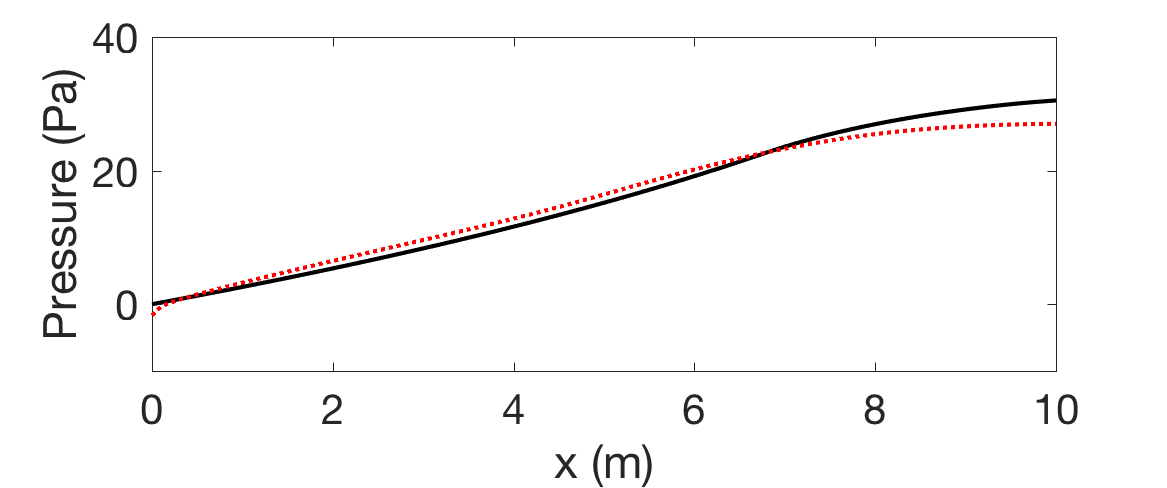}
\put(-5,40){(d)}
\end{overpic}
}
\caption{ Comparison of the simplified model and the $k$-$\epsilon$ model for asymmetric shear flow in a straight channel, where flow is from left to right. (a, b) Time-averaged velocity in the $x$ direction, $u$, where in the case of the simplified model, (a), we also overlay dashed lines indicating the width of the shear layer. (c, d) Comparison of time-averaged velocity and pressure profiles calculated using the $k$-$\epsilon$ model, in red, and the simplified model, in black (the simple model does not resolve the velocity profile in the wall boundary layers).   \label{asymmetric}}
\end{figure}

First, we compare our model to a steady $k$-$\epsilon$ turbulence model \citep{launder1974numerical} using the open-source software package \textit{OpenFoam} \citep{weller1998tensorial}. We choose a straight two-dimensional geometry of width $h=1\un{m}$ and length $30\un{m}$. {The two-dimensional geometry has a mesh of $20000$ elements, with $100$ elements in the $y$ direction and $200$ in the $x$ direction. Many other grid resolutions have been tested, and we find that solution for this grid resolution is well within the convergence regime.} We choose an inlet flow profile with $h_1(0)=h_2(0)$, $\delta(0)=0$ and $U_2(0)/U_1(0)=0.5$, where the speed of the faster plug region is $U_1(0)=1\un{m/s}$. Inlet conditions for the turbulence variables $k$ and $\epsilon$ are given by the free-stream boundary conditions \citep{schlichting1960boundary} $k=I^2\times3/2\lb u^2+v^2\rb$ and $\epsilon=0.09 k^{3/2}/\ell$, with turbulence intensity $I=10\%$ {(motivated by PIV calculations, which are discussed later)} and mixing length $\ell=0.1h$. If we use the channel width $h$ as a typical length scale, the faster plug region inlet speed $U_1(0)$ as a typical speed and choose the viscosity of water $\nu=10^{-6} \un{m^2/s}$, then the Reynolds number is $Re=10^6$. For comparison with the simple mathematical model, we use the Blasius relationship to estimate the friction factor, giving a value of $f=0.01$. 
For the spreading parameter $S_c$, we find that $S_c=0.18$ fits best with the CFD calculations, {which is larger than $S$ for free shear layers {($S=0.06-0.11$)} \citep{pope2000turbulent}. However, we find that this is consistent with later PIV experiments in two different three-dimensional geometries and at a different Reynolds number.}

In Figure \ref{asymmetric} we display colour plots of the streamwise velocity $u$ generated with both the simple model and the $k$-$\epsilon$ model. We also compare velocity profiles at equally spaced locations in the channel and the pressure profile, averaged across the width of the channel. 
We see that the shear layer grows across the channel, causing a rise in pressure due to the diffusion of momentum. 
{In our model, the growth of the shear layer is controlled by $S_c$ and the pressure is controlled by both $f$ and $S_c$. }
There is a strong correlation for both velocity and pressure (with an average error of $\sim 5\%$). 
Near the wall the velocity comparison is less accurate due to the fact that the model does not explicitly resolve the development of boundary layers. {However, it is apparent that the friction factor accurately represents the effect of the boundary layers on the pressure (wall drag) since there is a close comparison in Figure \ref{asymmetric} (d).} {We have also made comparisons with other CFD models on the same geometry, such as the $k$-$\omega$ Shear Stress Transport model (SST)} \citep{menter1992improved}{, and the results are similar.} {The close agreement between these CFD models is probably due to the fact that there are no strong adverse pressure gradients in the examples we study.}
{Furthermore, we have investigated longer channels and find that the simple model continues to show good comparison, eventually recovering a fully developed profile, though we do not display these results here. }

\begin{figure}
\centering
\includegraphics[width=0.5\textwidth]{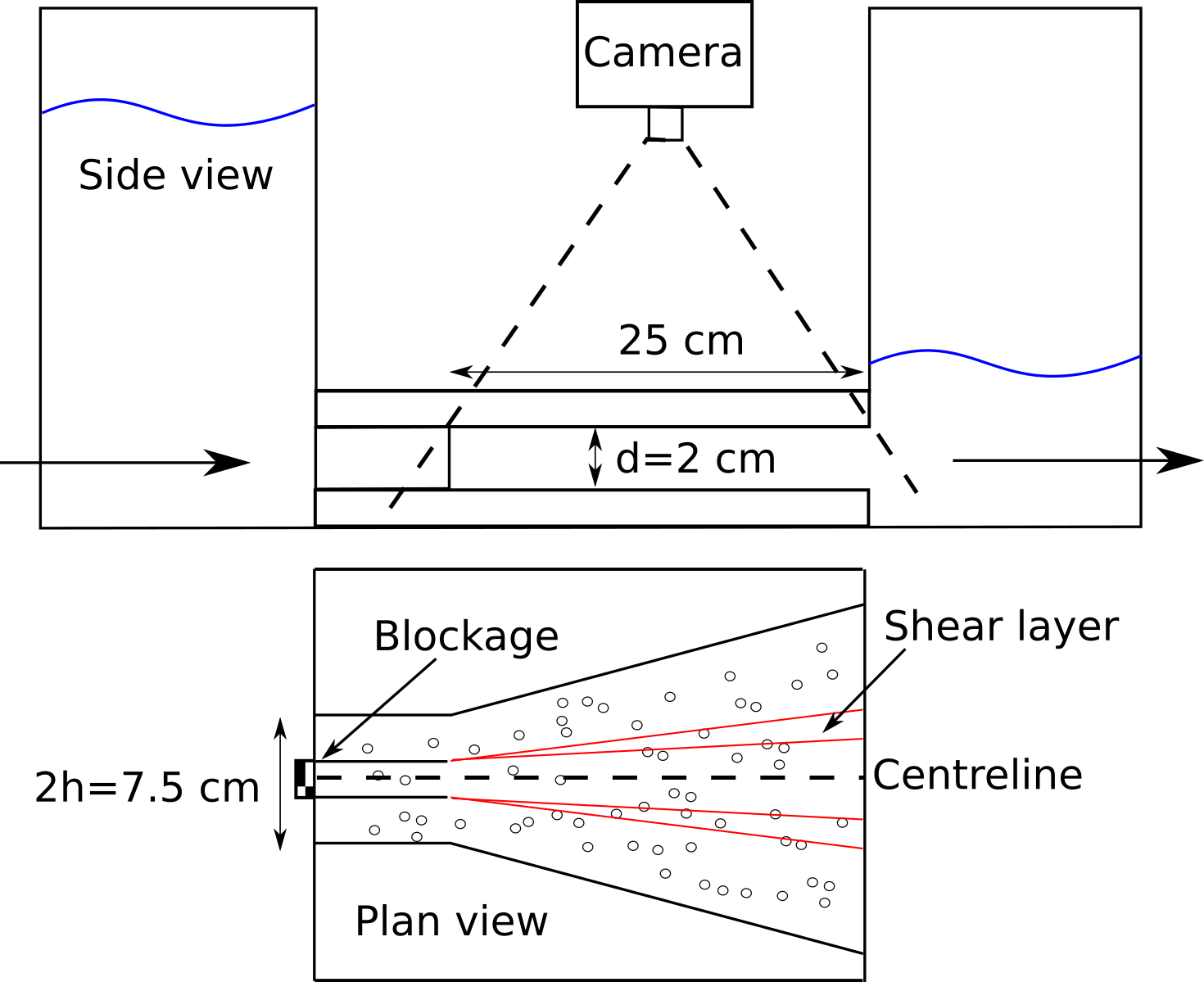}
\caption{Experimental setup, showing the configuration for a widening rectangular channel. A recirculating pump provides a constant flow rate. \label{expschem}}
\end{figure}

Next we validate our model by comparison with Particle Image Velocimetry (PIV) experiments.
In the experiments, water flows through an acrylic closed channel of rectangular cross-section at constant flow rate (see Figure \ref{expschem}).
At the channel inlet, three inflows merge to generate shear layers: two fast, outer flows, and a slow (partially-blocked) inner flow along the channel centre.
We use two different configurations. In the first, the walls of the channel are straight so that there is no flow dilation.
In the second, the walls of the channel widen at an angle of 5$^\circ$. The channel dimensions are $25\un{cm}\times 7.5\un{cm} \times 2\un{cm}$. If we take a typical length scale as {the hydraulic diameter 
$2\times \mathrm{width}\times\mathrm{depth}/( \mathrm{width}+\mathrm{depth})=3.2\un{cm}$}, and a typical velocity scale as the average speed of the fast outer flow measured by PIV ($50\un{cm/s}$), then the Reynolds number is $Re \approx10^4$. At the inlet the flow is observed to be turbulent, indicated by significant turbulent fluctuations (see Figure \ref{results2}) and a turbulent intensity of $I=\sqrt{2/3 k/\lb u^2+v^2\rb}\approx10\%$ \citep{schlichting1960boundary}, as measured by PIV. {Another possible method to measure $I$ is by using hot-wire anemometry, though this was not investigated for this particular study. }

We perform PIV by adding a dilute suspension of neutrally-buoyant, pliolite, tracer particles to the flow \citep{sveen2005dynamic,dalziel2007simultaneous}, and shining a pulsed \textit{Nd:YAG} laser sheet through the channel side wall at its halfway height.
A synchronised high-speed camera takes images at $1\un{ms}$ intervals controlled with the \textit{TSI Insight 4G} computer package \citep{TSI}. 
\textit{Matlab}'s package \textit{PIVlab} \citep{guide1998mathworks,thielicke2014pivlab} then allows us to extract time-averaged properties of the flow such as the streamwise and transverse velocities $(u,v)$ (see video in Supplementary Materials) and the turbulent kinetic energy $k$ \citep{schlichting1960boundary, pope2000turbulent}.
Simultaneously we measure time-averaged pressures along the channel using regularly-spaced, open-tube-manometer pressure tappings.

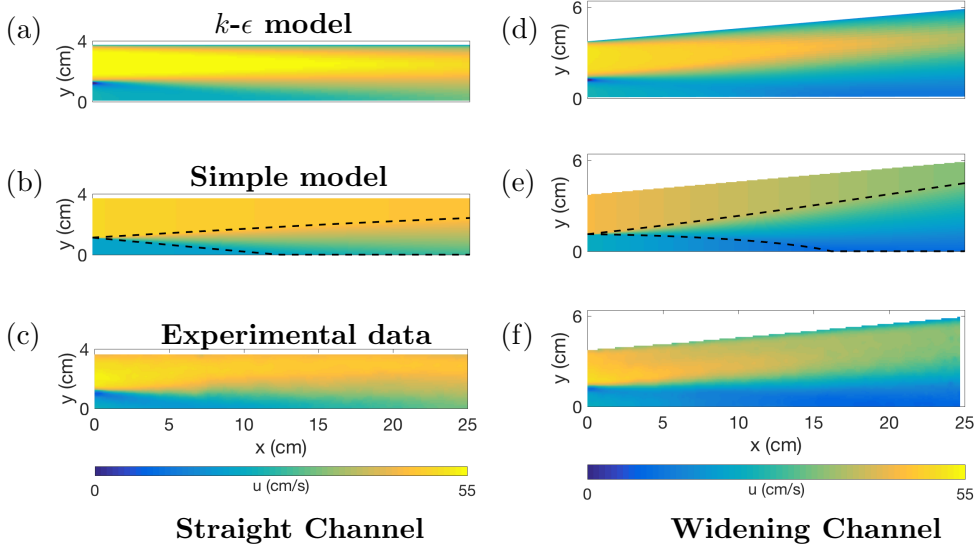
\begin{figure}
\centering
\subfloat{
\begin{overpic}[width=0.45\textwidth]{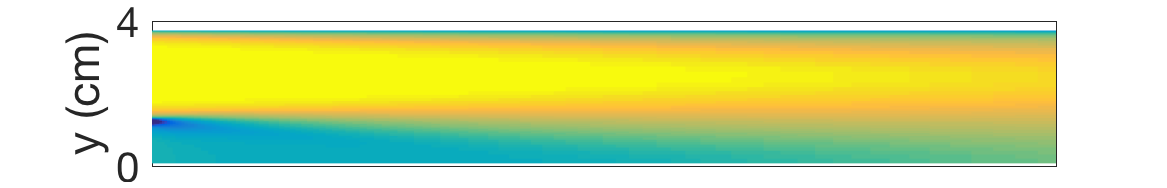}
\put(-5,15){(a)}
\put(38,16){\bf $k$-$\epsilon$ model}
\end{overpic}
\begin{overpic}[width=0.45\textwidth]{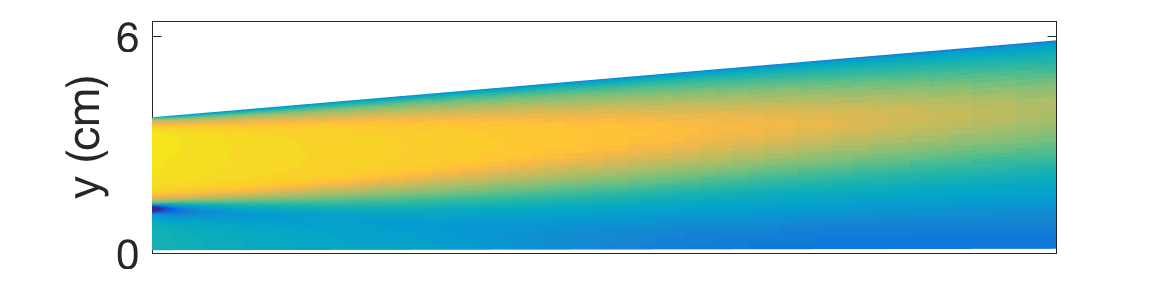}
\put(-5,15){(d)}
\end{overpic}
}
\vspace{0.1cm}\\
\subfloat{
\begin{overpic}[width=0.45\textwidth]{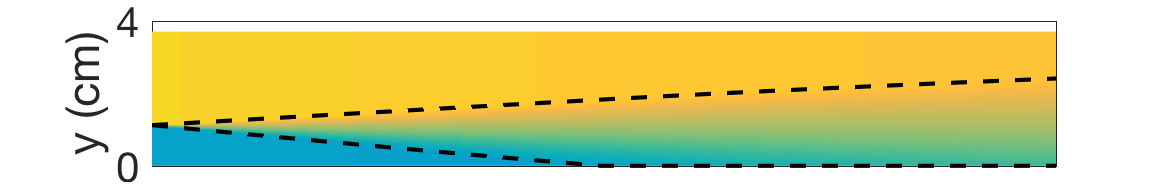}
\put(-5,15){(b)}
\put(33,16){\bf Simple model}
\end{overpic}
\begin{overpic}[width=0.45\textwidth]{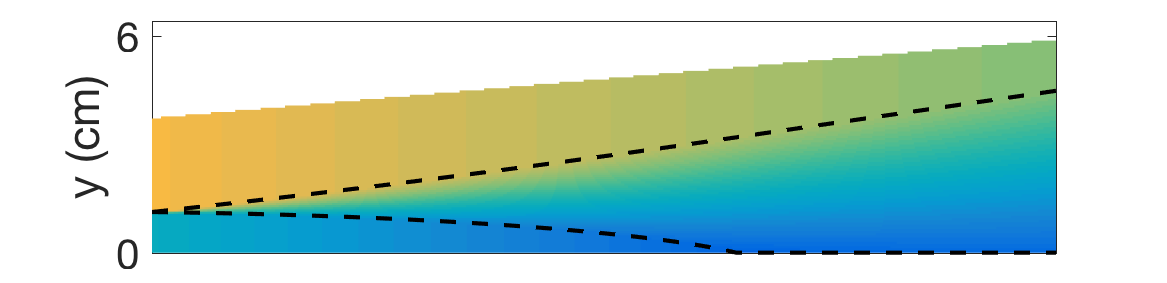}
\put(-5,15){(e)}
\end{overpic}
}
\vspace{0.1cm}\\
\subfloat{
\begin{overpic}[width=0.45\textwidth]{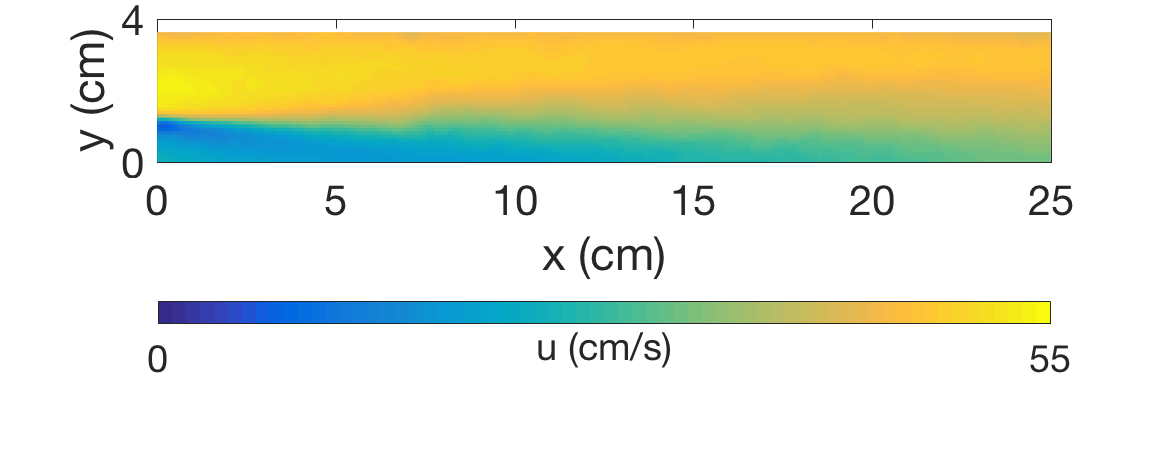}
\put(-5,40){(c)}
\put(27,40){\bf Experimental data}
\put(30,0){\bf Straight Channel}
\end{overpic}
\begin{overpic}[width=0.45\textwidth]{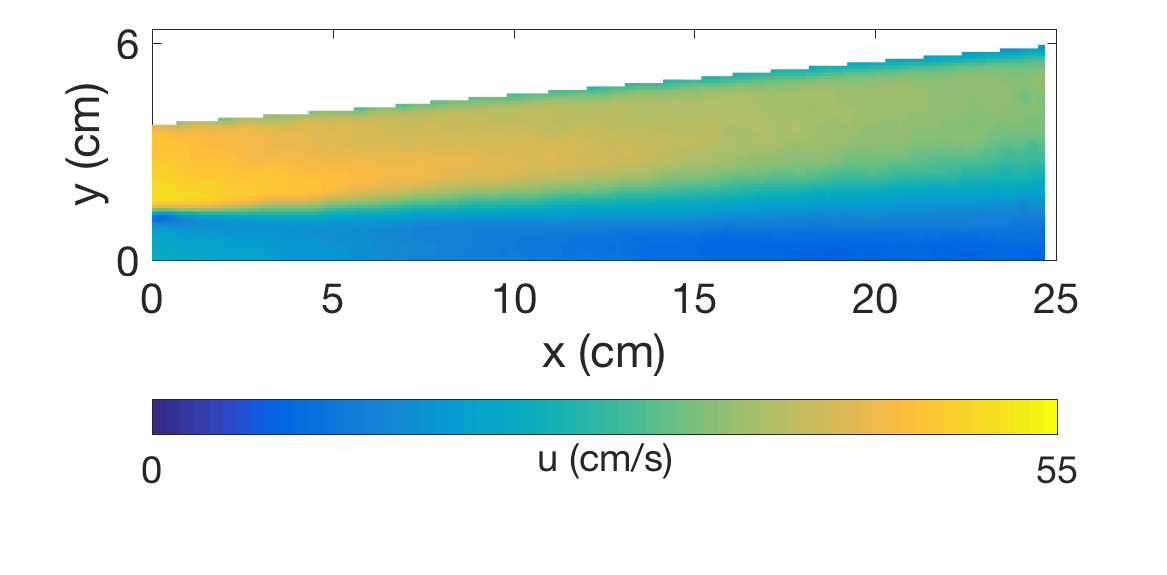}
\put(-5,40){(f)}
\put(30,0){\bf Widening Channel}
\end{overpic}
}
\caption{Comparison of $k$-$\epsilon$ model, simple model and experimental data for straight channel (a, b, c), and widening channel (d, e, f). Flow is from left to right and is symmetric about the channel centreline so we only show half of the channel. Colour plots show time-averaged velocity in the $x$ direction, $u$. In the case of the simplified model, we also overlay dashed lines representing the width of the shear layer. \label{results}}
\end{figure}

\begin{figure}
\centering
\subfloat{
\begin{overpic}[width=0.45\textwidth]{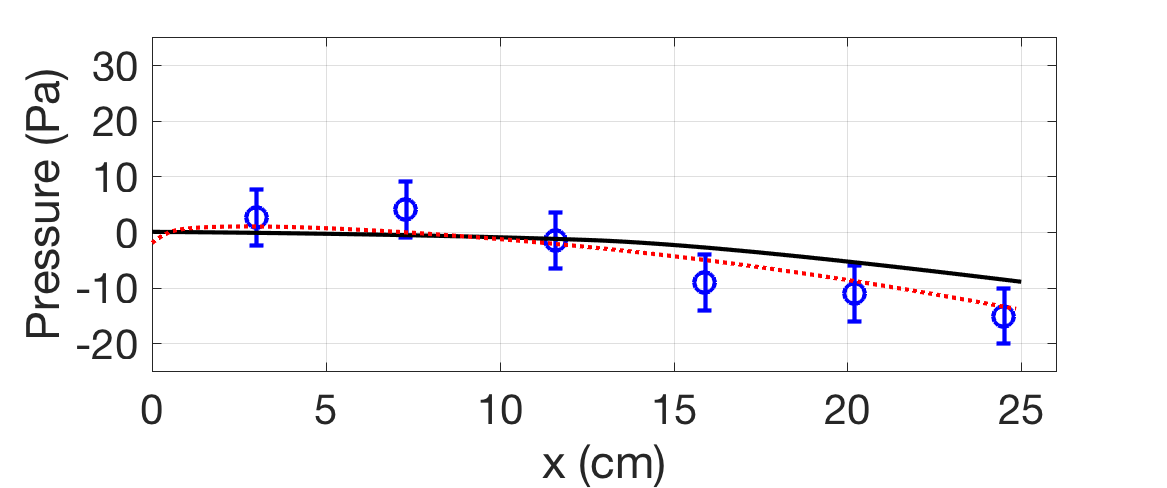}
\put(-5,35){(a)}
\end{overpic}
\begin{overpic}[width=0.45\textwidth]{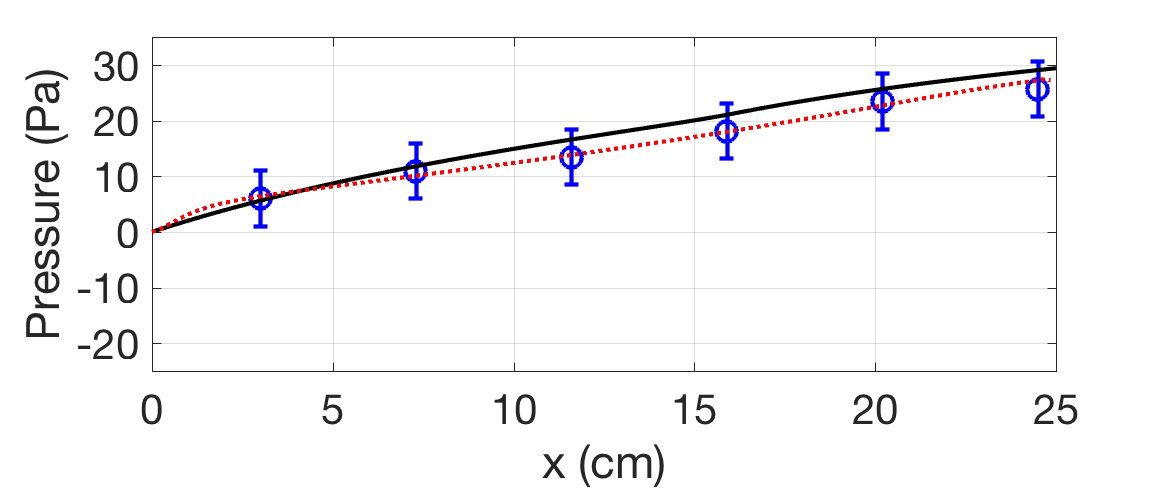}
\put(-5,35){(d)}
\end{overpic}
}\\
\subfloat{
\begin{overpic}[width=0.45\textwidth]{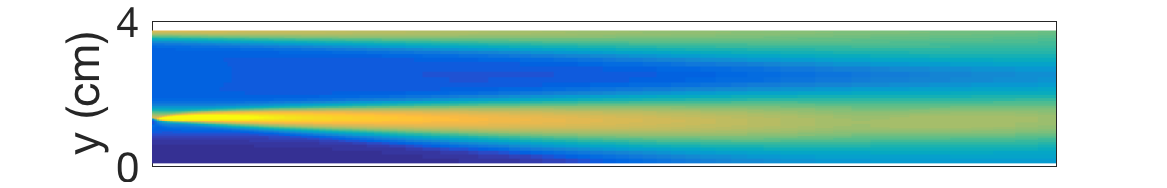}
\put(-5,15){(b)}
\put(38,16){\bf $k$-$\epsilon$ model}
\end{overpic}
\begin{overpic}[width=0.45\textwidth]{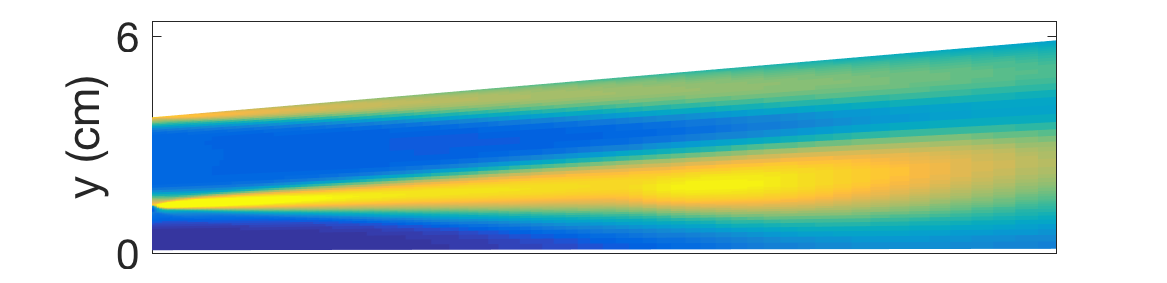}
\put(-5,20){(e)}
\end{overpic}
}\\
\subfloat{
\begin{overpic}[width=0.45\textwidth]{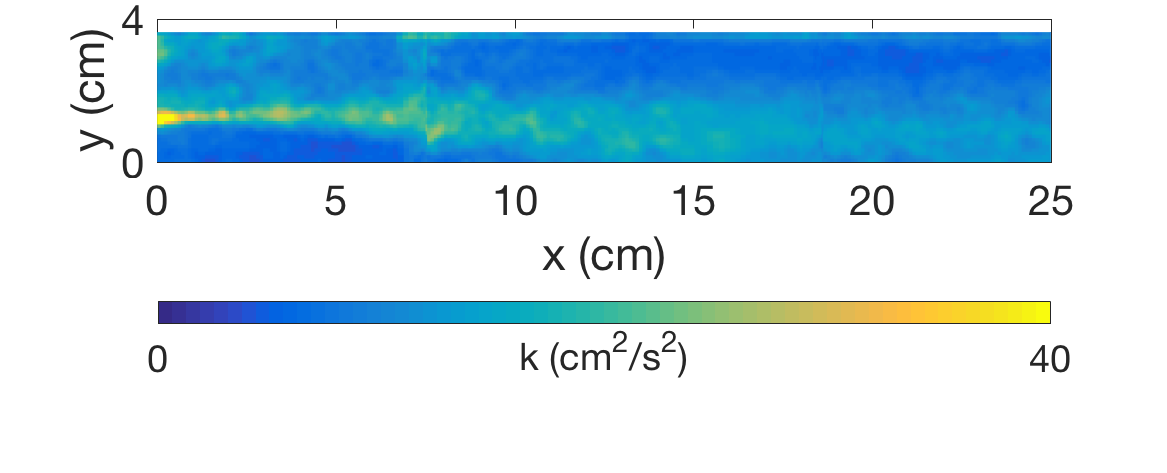}
\put(-5,40){(c)}
\put(27,40){\bf Experimental data}
\put(30,0){\bf Straight Channel}
\end{overpic}
\begin{overpic}[width=0.45\textwidth]{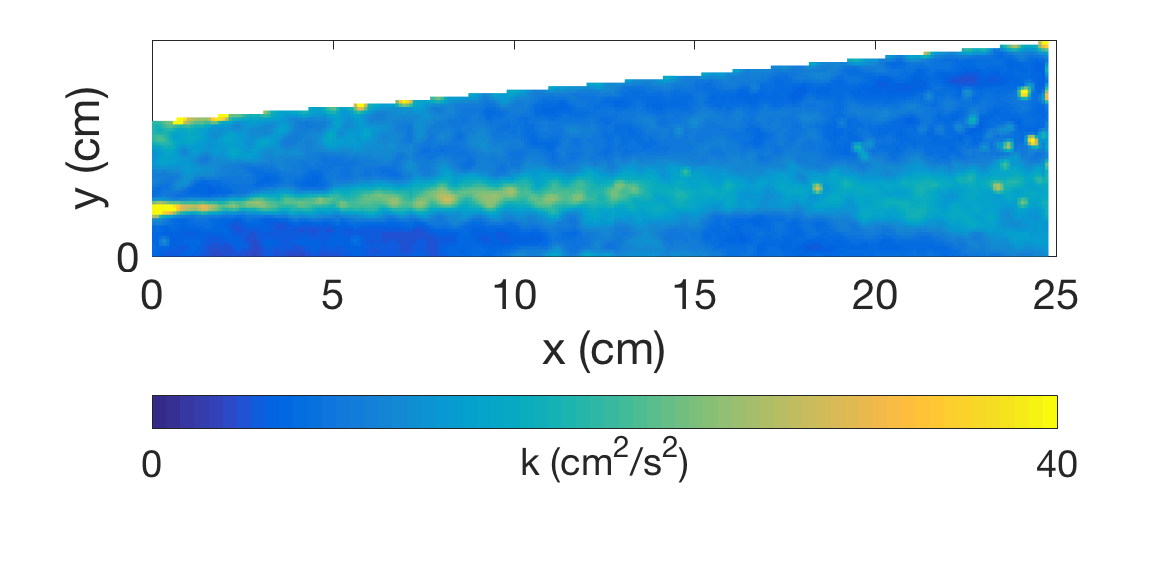}
\put(-5,45){(f)}
\put(30,0){\bf Widening Channel}
\end{overpic}
}
\caption{Comparison of $k$-$\epsilon$, simple model and experimental data, continued from Figure \ref{results}. In (a, d) streamwise pressure profiles are shown, with blue data points representing experimental data and red and black lines representing calculations from the $k$-$\epsilon$ model and the simplified model, respectively. Error margins correspond to an accuracy of $0.5\un{mm}$ in human eye measurements. In (b, c, e, f) time-averaged turbulent kinetic energy $k$ is shown, calculated using the $k$-$\epsilon$ model and PIV measurements. \label{results2}}
\end{figure}
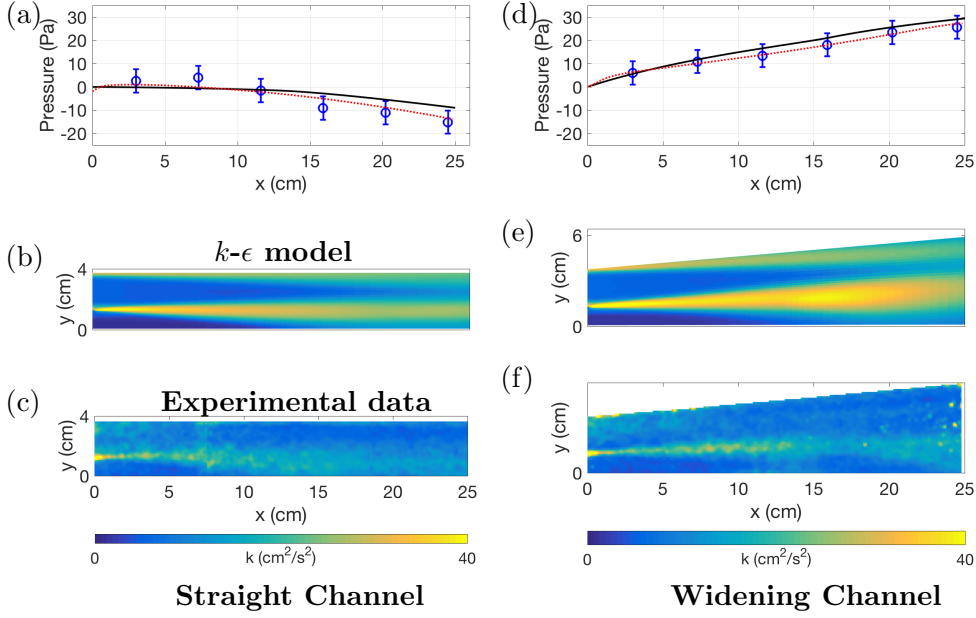

The experimental results are shown in Figure \ref{results} (c, f) for nominally the same inflow velocities in both straight and widening channels.
We plot colour maps of the local, time-averaged flow speed $u$.
These clearly show the flow mixing and, in the second case, the overall flow deceleration in the widening channel.
In the straight channel case, mixing occurs over the whole channel length, with the faster streams slowing down and the slower stream speeding up.
In the case of the widening channel, the expansion causes the flows to slow down at different rates, accentuating the non-uniform flow profile. At $x=25\un{cm}$ the slower central stream almost stops entirely. 
The location of the shear layer is more clearly shown by the plots of turbulent kinetic energy in Figure \ref{results2} (c, f) where we see the expansion of the layer, as well as some evidence of wall friction at the channel's side walls.

As before, CFD modelling is performed with a steady $k$-$\epsilon$ turbulence model \citep{launder1974numerical} on the same three-dimensional geometry as the experiment. Inlet conditions were modeled slightly further upstream at $x=-10\un{cm}$, corresponding to the three inflows in the experiment (see Figure \ref{expschem}), with plug flow profiles for each of the three inflows such that the mass flux is consistent with PIV measurements. {The three-dimensional geometry has a mesh of $65000$ elements, with $100$ elements in the $x$ direction, $65$ in the $y$ direction and $10$ in the $z$ direction. As before, we have tested for convergence and we find that this grid resolution is well within the convergence regime.} Inlet conditions for the turbulence variables $k$ and $\epsilon$ are given by the free-stream boundary conditions, as before, with turbulence intensity $I=10\%$ and mixing length $\ell=1\un{cm}$ (same order of magnitude of the width of each of the inlet flows).

For comparison with the simple mathematical model, we used the PIV-measured speeds of the streams at the inlet $x=0$ as initial conditions for $U_1$ and $U_2$.
{Since the channel is symmetric, we only consider half of the channel in our simplified model $0\leq y \leq h$. The governing equations of the model (\ref{shear}) - (\ref{mass_moment}) apply in the half-channel, except with modified stress terms in Equation (\ref{mass_moment}). Due to symmetry, there is zero stress at $y=0$, such that $\tau_{w_0}=0$.
Furthermore, in addition to drag from the side walls, we account for wall drag on the lid and base of the channel by adding an extra stress term to (\ref{mass_moment}) of the form
\beq
\tau_{w_d}=  \frac{2}{d}\int^h_0 -\frac{1}{8}f\rho u^2\, dy  ,\label{3dstress}
\eeq
where $d$ is the depth of the channel in the third dimension {(see Figure \ref{expschem})}. The drag from the top and bottom walls also requires a modification of Bernoulli's equations (\ref{bernoulli}) that apply in the two plug regions, which now become
\beq
\frac{d}{dx}\lb p+\frac{1}{2}\rho U_i^2\rb = -\frac{2}{d}\lb \frac{1}{8}f\rho U_i^2\rb,  \quad i=1,2.\label{bernoulli3d}
\eeq}
In Appendix \ref{appB}, we show how (\ref{mass_moment}), (\ref{3dstress}) and (\ref{bernoulli3d}) can be derived from the turbulent boundary layer equations.
We use the Blasius relationship to estimate the friction factor, giving a value of $f=0.03$. 
{
We find that the value of the spreading parameter $S_c$ that fits best with our data is $S_c=0.18$, which is identical to the two-dimensional case, indicating that the model is consistent.}
The simple model, $k$-$\epsilon$ model and experiments show good agreement.
For example, the streamwise velocities $u$ are compared in Figure \ref{results}. 
Figure \ref{results2} also shows a comparison of the pressure change (relative to the inlet pressure) along the channel centreline. 
{Dominant features of the flow, such as maximum and minimum velocities and pressure variation, are captured accurately by the simple model and in some cases more accurately than the $k$-$\epsilon$ model. For example, the $k$-$\epsilon$ model over-predicts the maximum velocity in both the straight and widening channel cases.}

{After comparison with both CFD and experimental data in several geometries and at two different Reynolds numbers, we have shown that the simple model presented here is robust and has strong capabilities in predicting the dominant features of the flow and pressure.} 
It can therefore serve as a useful, low-computational-cost tool in modelling and optimising processes that involve confined shear layers, such as diffusers and nozzles. Furthermore, although we do not discuss it here, the model can easily be extended to axisymmetric flows by following the derivation in Appendix \ref{appB} except with the cylindrical polar form of the turbulent boundary layer equations.

\section{Diffuser shape optimisation}

\begin{figure}
\centering
\subfloat{
\begin{overpic}[width=0.45\textwidth]{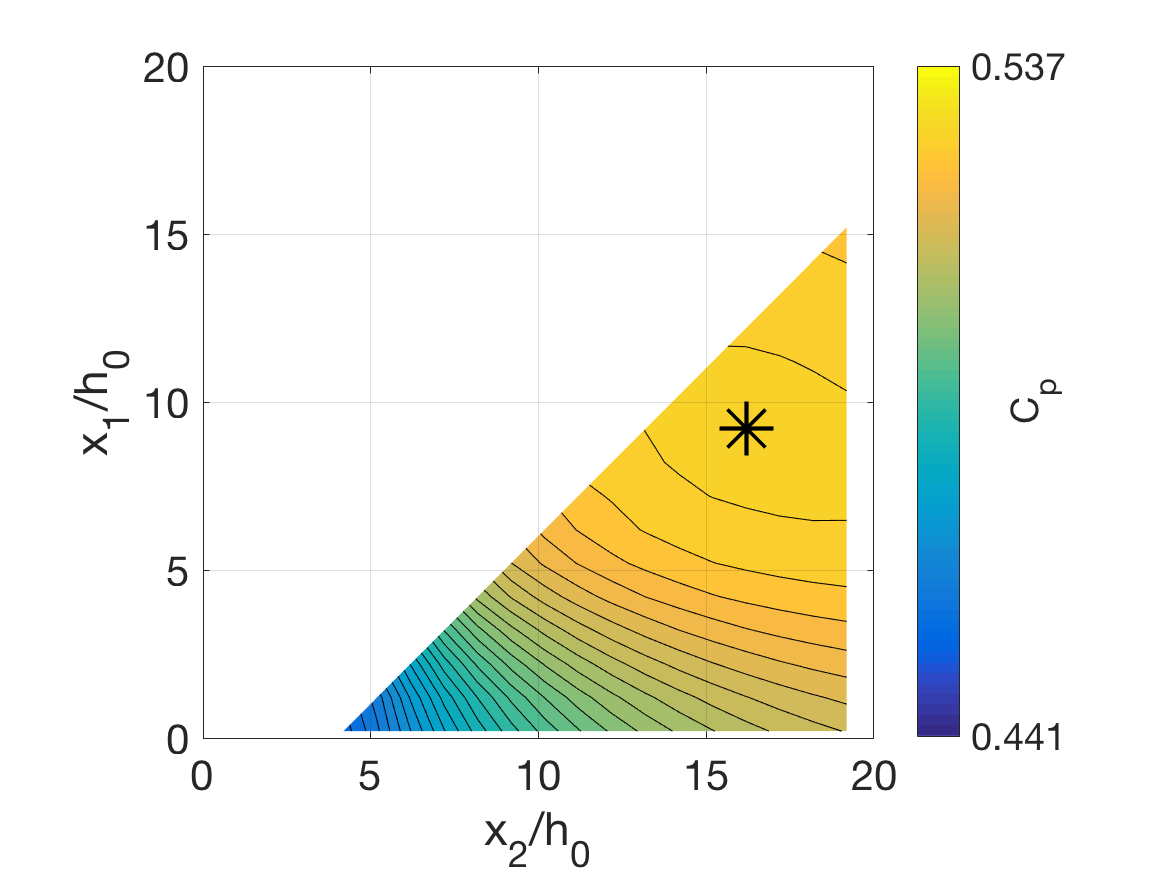}
\put(-5,65){(a)}
\put(22,60){\bf Simple model}
\end{overpic}
}
\subfloat{
\subfloat{
\begin{overpic}[width=0.45\textwidth]{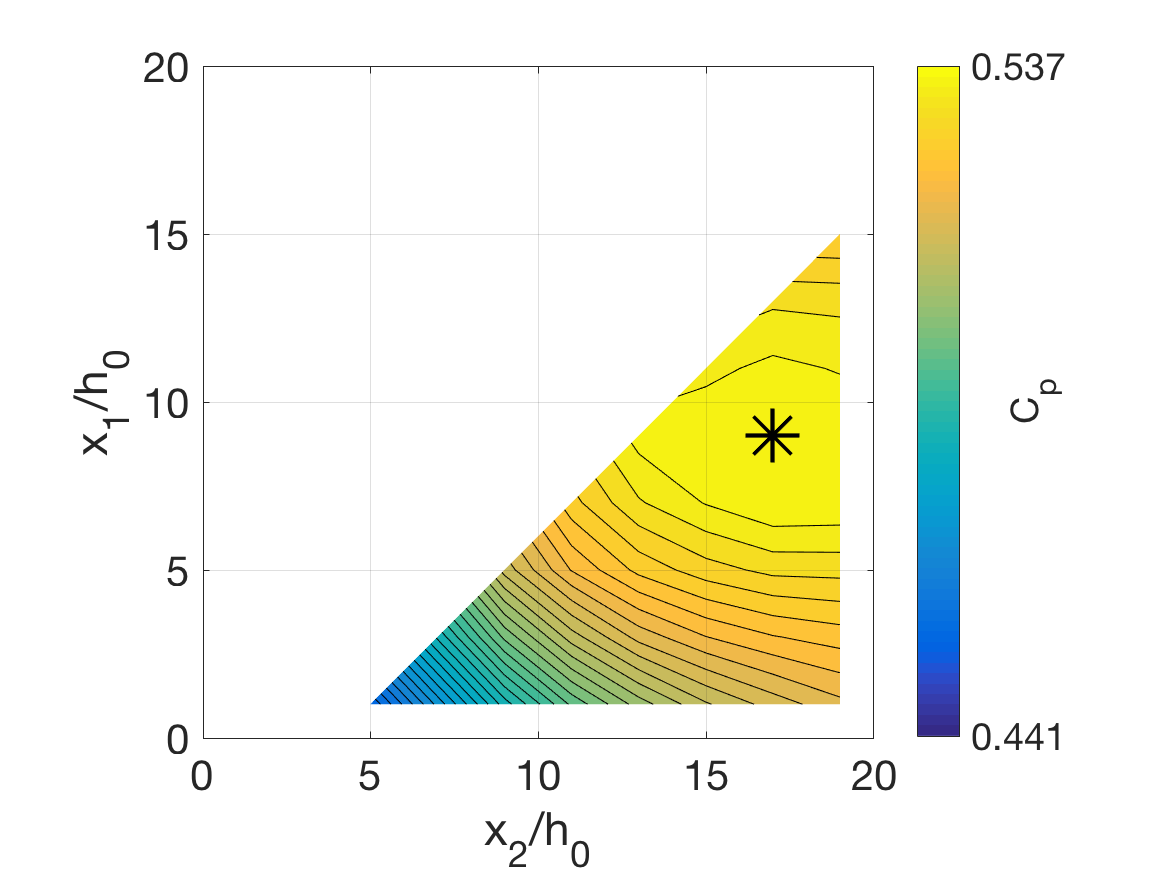}
\put(-5,65){(b)}
\put(25,60){\bf $k$-$\epsilon$ model}
\end{overpic}
}
}\\
\begin{overpic}[width=0.45\textwidth]{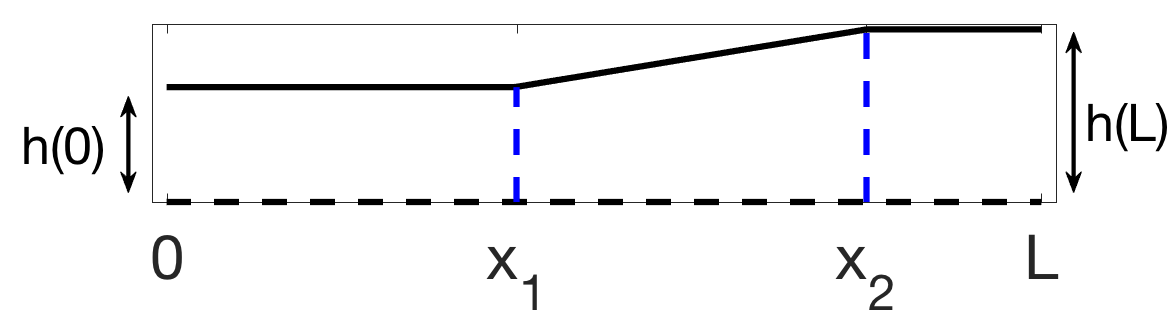}
\put(-5,22){(c)}
\end{overpic}
\caption{Optimum design of a diffuser with a non-uniform inlet flow profile. The diffuser, composed of two straight sections and widening section, is $20$ half-widths long $L/h(0)=20$, and has area ratio $h(L)/h(0)=1.5$. The two parameters $x_1$ and $x_2$ define the split between straight and widening sections. In (a, b) contour plots of the pressure recovery coefficient (\ref{C_p}) are computed using the simplified model and the $k$-$\epsilon$ model, respectively, using the same colour scale. {The optimum point for (a) is  $x_1/h(0)=9.2$ and $x_2/h(0)=16.2$, giving $C_p=0.528$, compared to (b), which has $x_1/h(0)=9$ and $x_2/h(0)=17$, giving $C_p=0.537$. } \label{optimal}}
\end{figure}

As an application of the simplified model, we study the optimal design of a two-dimensional diffuser with a non-uniform inlet flow profile. We restrict our attention to a particular class of shapes that are characterised by three linear sections: a straight section for $0\leq x\leq x_1$, followed by a widening section of constant opening angle for $x_1< x\leq x_2$, followed by another straight section for $x_2< x\leq L$, where $L$ is the total length of the profile (see Figure \ref{optimal}). This class of diffuser shapes is both realistic and has the advantage that it can be defined by a very small number of parameters.
We measure diffuser performance by its mass-averaged pressure recovery coefficient -- a measure of the pressure gained in the diffuser, relative to the kinetic energy flux at the inlet \citep{filipenco1998effects},
\begin{equation}
C_p=\frac{ \int_0^h u p \,dy |_{x=L} - \int_0^h u p \,dy |_{x=0}}{\int_0^h \frac{1}{2}\rho u^3 \,dy |_{x=0} }.\label{C_p}
\end{equation}
Note that $C_p\in[-\infty,1]$, with $C_p=1$ when all the dynamic pressure of the inlet flow is converted into static pressure. For a fixed finite area ratio $h(L)/h(0)$ and inlet flow profile, there is a maximum possible pressure recovery which is less than $1$ \citep{blevins1984applied}. In the case of uniform inviscid flow this ideal limit is $C_{p_i}=1-\lb h(0)/h(L)\rb^2$ but for non-uniform flow, it is not known what the corresponding limit is  \citep{blevins1984applied}.

For the purpose of this optimisation, we consider a two-dimensional symmetric diffuser which is $20$ half-widths long $L/h(0)=20$ and has area ratio $h(L)/h(0)=1.5$. The inlet profile is taken to be symmetric, as in the experiments, except with $U_{2}(0)/U_{1}(0)=0.4$ and $h_{2}(0)=h_1(0)$. This setup is motivated by a hydropower application in which the central slower stream emerges from a turbine. In Figure \ref{optimal} we show a contour plot of $C_p$ over all the possible diffuser shapes considered, calculated using both the simplified model and the $k$-$\epsilon$ model. {In our simple model, we use the same value of the spreading parameter $S_c=0.18$ and the friction factor $f=0.01$, as in Figure 2. }
We restrict the parameter space such that the opening angle of the widening section is less than $7^\circ $. This is because separation starts to occur for angles larger than around this value \citep{blevins1984applied} and the simple model is incapable of accounting for separation effects. {According to the simplified model, there is an optimum at $x_1/h(0)=9.2$ and $x_2/h(0)=16.2$, giving an optimal pressure recovery of $C_p=0.528$. Note that this is $95\%$ of the ideal inviscid value for uniform flow $C_{p_i}=0.556$. According to the $k$-$\epsilon$ model, the optimum is almost identical, with $x_1/h(0)=9$ and $x_2/h(0)=17$, and an optimal pressure recovery of $C_p=0.537$.} The optimum is striking a balance between not widening too soon, which would exacerbate the non-uniform flow, and not staying narrow for too long, which would increase wall friction losses.

\section{Summary and concluding remarks}

We have developed a simple model for turbulent shear layers in confined geometries that shows good agreement with PIV experiments and CFD {for a variety of channel shapes and at two different Reynolds numbers. 
The model depends on two parameters, the friction factor $f$, which we find is well estimated by the Blasius relationship \cite{blasius1913ahnlichkeitsgesetz}, and the spreading parameter $S_c$ which is fitted to the CFD and experimental data, giving a consistent value of $S_c=0.18$. {It is possible that the value of $S_c$ is not universal for all geometries and Reynolds numbers. However, although we do not display the results here, we have compared our simple model to calculations from a $k$-$\epsilon$ model and a $k$-$\omega$ SST} \cite{menter1992improved} {model, in a variety of other geometries and at Reynolds numbers between $Re=10^4$ and $Re=10^6$, and we find that $S_c=0.18$ is consistent.}

We have also tried allowing $f$ to vary with $x$, as the average velocity (and hence the Reynolds number $Re$) in the channel changes, but we find that this makes little difference to the results. This can be explained by the fact that $f$ depends weakly on $Re$ \cite{blasius1913ahnlichkeitsgesetz}. Furthermore, in the expanding geometry (Figure \ref{results} (d, e, f)) the hydrodynamic diameter increases whilst the average velocity decreases, such that $Re$ stays approximately constant. }

The model's simplicity makes it ideal for simulation and design purposes for a range of problems from diffuser design to urban and geophysical flows.
In particular, given its low computational cost, it can be used to perform shape optimisation to maximise pressure recovery for a given inlet flow profile.

{
For future work, we aim to extend the model to account for boundary layer development. In this case, the model for the boundary layers would replace the friction factor in calculating the wall stress. However, we have seen that the friction factor and the Blasius relationship perform well in a variety of flow situations.
}

\section*{Acknowledgements}
This publication is based on work supported by the EPSRC Centre for Doctoral Training in Industrially Focused Mathematical Modelling (EP/L015803/1) in collaboration with VerdErg Renewable Energy Limited and inspired by their Venturi-Enhanced Turbine Technology for low-head hydropower. Experimental work was supported by the John Fell Fund (Oxford University Press). A.A. Castrejon-Pita was supported by a Royal Society URF. \\



\appendix
\section{}\label{appA}
In Section \ref{model} we motivate Equation (\ref{shear}) by analogy with free shear layers. Here we give an alternative derivation based upon an entrainment assumption.

Consider the two dimensional velocity profile given by Equation (\ref{piecewise}). We hypothesise that the shear layer entrains fluid from each of the plug regions. Similarly to a Morton-Taylor-Turner jet \citep{morton1956turbulent}, we assume that the rate of entrainment is proportional to the difference in speeds between the two plug regions $E=1/2S_c(U_1-U_2)$, where $S_c$ is an entrainment constant. Thus, the conservation of mass equations for each of the plug regions and the shear layer, are
\begin{align}
\frac{d}{d x}\lb U_1h_1 \rb&=-\frac{1}{2}S_c(U_1-U_2),\label{mass1}\\
\frac{d}{d x}\lb U_2h_2 \rb&=-\frac{1}{2}S_c(U_1-U_2),\label{mass2}\\
\frac{d}{d x}\lb \frac{1}{2}\lb U_1+U_2\rb\delta \rb&=S_c(U_1-U_2)\label{mass3}.
\end{align}
In the case of a free shear layer, where $U_1$ and $U_2$ are constant, Equation (\ref{mass3}) is equivalent to Equation (\ref{delta}). This is expected because Equation (\ref{delta}) comes directly from the theory of free shear layers \citep{schlichting1960boundary}. Using the definition of the shear rate within the shear layer, $\epsilon_y=(U_1-U_2)/\delta$, and Equation (\ref{mass3}), we find that
\beq
\frac{U_1+U_2}{2}\frac{d\epsilon_y}{d x}=-S_c\epsilon_y^2+\frac{1}{\delta}\lb U_1\frac{d U_1}{d x}-U_2\frac{d U_2}{d x}\rb.\label{shearderiv1}
\eeq
If we assume that Bernoulli's equation (\ref{bernoulli}) holds in each plug region then we have
\beq
\frac{d}{d x}\lb \frac{1}{2}U_1^2-\frac{1}{2}U_2^2\rb=0,\label{shearderiv2}
\eeq
so (\ref{shearderiv1}) becomes
\beq
\frac{U_1+U_2}{2}\frac{d\epsilon_y}{d x}=-S_c\epsilon_y^2,
\eeq
which is the shear layer Equation (\ref{shear}).

\section{}\label{appB}
Consider incompressible flow in a three dimensional channel defined by $0\leq x \leq L$, $0\leq y\leq h(x)$, $0\leq z\leq d$, where $h\ll L$ and  $d\ll L$. {Because the domain is long and thin, the boundary layer approximation to the Navier-Stokes equations is appropriate for modelling the flow in the whole domain \citep{schlichting1960boundary}.} The three-dimensional turbulent boundary layer equations for the time-averaged velocities $(u,v,w)$ and pressure $p$ are
\begin{align}
\frac{\partial u}{\partial x}+\frac{\partial v}{\partial y} +\frac{\partial w}{\partial z} &=0,\label{blmass}\\
u\frac{\partial u}{\partial x} +v\frac{\partial u}{\partial y} +w\frac{\partial u}{\partial z} &=-\frac{1}{\rho}\frac{\partial p}{\partial x}+\frac{\partial}{\partial y}\lb \lb \nu+\nu_t\rb \frac{\partial u}{\partial y}\rb +\frac{\partial}{\partial z}\lb \lb \nu+\nu_t\rb \frac{\partial u}{\partial z}\rb,\label{bl1}\\
0&=-\frac{\partial p}{\partial y},\label{bl2}\\
0&=-\frac{\partial p}{\partial z},\label{bl3}
\end{align}
where $\rho$ is the density, $\nu$ is the viscosity and $\nu_t$ is the eddy viscosity. Firstly note that because of (\ref{bl2}) and (\ref{bl3}) the pressure varies in the $x$ direction alone $p=p(x)$. We impose the no-slip condition on the walls at $z= 0,\,d$ and $y= 0,\,h(x)$, such that
\begin{align}
u=v=w&=0,\quad \mathrm{on}\quad y= 0,\,h(x),\label{bc1}\\
u=v=w&=0,\quad \mathrm{on}\quad z= 0,\,d.\label{bc2}
\end{align}
Now we integrate equations (\ref{blmass}) and (\ref{bl1}) across the channel in the $y$ and $z$ directions, using boundary conditions (\ref{bc1}) -  (\ref{bc2}),  to give
\begin{align}
\frac{d}{d x}\lb \int_0^{d} \int_0^{h(x)} \rho u\, dy\, dz\rb &= 0,\label{avmass}\\
\frac{d}{d x}\lb \int_0^{d} \int_0^{h(x)} \rho u^2\, dy\, dz\rb &= -d h(x) \frac{d p}{d x} + d \lb \tau_{w_0} + \tau_{w_h}\rb + h\lb \tau_{w_t} + \tau_{w_b}\rb,\label{avmom}
\end{align}
where the wall stress terms are
\begin{align}
\tau_{w_0}&=\frac{1}{d}\int_0^d\left[ \rho \lb \nu+\nu_t\rb \frac{\partial u}{\partial y}\right]_{y=0}\,dz, \label{stress1}\\
\tau_{w_h}&= \frac{1}{d}\int_0^d \left[ \rho \lb \nu+\nu_t\rb \frac{\partial u}{\partial y}\right]_{y=h}\,dz,\label{stress2}\\
\tau_{w_t}&=\frac{1}{h}\int_0^h\left[ \rho \lb \nu+\nu_t\rb \frac{\partial u}{\partial z}\right]_{z=d}\,dy, \label{stress3}\\
\tau_{w_b}&= \frac{1}{h}\int_0^h \left[ \rho \lb \nu+\nu_t\rb \frac{\partial u}{\partial z}\right]_{z=0}\,dy.\label{stress4}
\end{align}
We now approximate the solutions to these equations by decomposing the flow into an `outer flow', comprising two plugs and a shear layer, as described in the main text, and narrow boundary layers adjacent to the walls that allow the no-slip conditions to be satisfied.  The latter are not resolved explicitly but their effect is parameterised by writing the wall stress terms (\ref{stress1}) - (\ref{stress4}) in terms of the outer velocity near the wall and a friction factor.  Thus, the bulk of the flow profile is taken to be that given by (\ref{piecewise}).
The wall stress terms on the right hand side of (\ref{avmom}) are replaced with
\beq
d \lb \tau_{w_0} + \tau_{w_h}\rb + h\lb \tau_{w_t} + \tau_{w_b}\rb=-\frac{1}{8}df\rho\lb U_1^2+U_2^2\rb + 2\int_0^{h} -\frac{1}{8}f\rho u^2 \, dy,\label{3dstress2}
\eeq 
where $f$ is an empirical friction factor. With these approximations, equations (\ref{avmass}) and (\ref{avmom}) rearrange to give
\begin{align}
\frac{d}{d x}\lb \int_0^{h(x)} \rho u\, dy\rb &= 0,\label{m1}\\
\frac{d}{d x}\lb \int_0^{h(x)} \rho u^2\, dy\rb +  h(x) \frac{d p}{d x} &=  -\frac{1}{8}f\rho\lb U_1^2+U_2^2\rb+\frac{2}{d}\int_0^{h}  -\frac{1}{8}f\rho u^2 \, dy.\label{m2}
\end{align}
Equations (\ref{3dstress2}) - (\ref{m2}) are identical to equations (\ref{mass}), (\ref{mass_moment}) and (\ref{3dstress}). It should be noted that if we remove the $z$ dimension (let $d\rightarrow \infty$), the final term in (\ref{m2}) vanishes and we are left with the two-dimensional form of the equations, as expected.

To derive Equation (\ref{bernoulli3d}) for the plug regions, we integrate (\ref{bl1}) over the depth of the channel, still assuming that it is approximately independent of both $y$ and $z$ in these regions, except in narrow boundary layers near the top and bottom. The result is 
\beq
\frac{d}{d x} \lb p+ \frac{1}{2}\rho U_i^2 \rb = \frac{1}{d}\left[\rho \lb \nu +\nu_t\rb \frac{\partial u}{\partial z}\right]^d_0 = -\frac{2}{d}\lb \frac{1}{8} f\rho U_i^2 \rb, \quad \mathrm{for}\quad i=1,2,
\eeq
where we have made use of the same friction factor parameterisation of the wall stress as above.

\bibliographystyle{tfnlm}
\bibliography{testbib.bib}

\end{document}